\documentclass[aps,prd,twocolumn]{revtex4}
\usepackage{graphicx}

\def\beq{\begin{equation}}
\def\eeq{\end{equation}}

\def\H{{\cal H}}
\def\N{{\cal N}}
\def\v{v}

\begin{document}

\title{Evolution of gravitational waves in Randall-Sundrum cosmology}

\author{Richard Easther$^1$, David Langlois$^2$, Roy Maartens$^3$,
David Wands$^{3,4}$}

\affiliation{~}

\affiliation{$^1$ISCAP, Columbia University, Mailcode 5247, New
York~NY~10027, USA}

\affiliation{$^2$Institut d'Astrophysique de Paris, CNRS,
75014~Paris, France}

\affiliation{$^3$Institute of Cosmology \& Gravitation, University
of Portsmouth, Portsmouth~PO1~2EG, UK}

\affiliation{$^4$Kavli Institute for Theoretical Physics, Kohn Hall, UC Santa Barbara, Santa Barbara~CA~93106, USA}

\begin{abstract}

We investigate the evolution of gravitational wave perturbations about a brane cosmology embedded in a five-dimensional anti-de Sitter bulk. 
During slow-roll inflation in a Randall-Sundrum brane-world, the
zero mode of the 5-dimensional graviton is generated, while the
massive modes remain in their vacuum state. When the zero mode
re-enters the Hubble radius during radiation domination, massive
modes are generated. We show that modes decouple in the low-energy/near-brane limit and develop perturbative techniques to calculate
the mode-mixing at finite energy.

\end{abstract}

\maketitle

\section{Introduction}

The Randall-Sundrum (RS) scenario~\cite{RS,RSII} is a simple and
novel way of realizing the idea that our observable universe could
be a 4-brane embedded in a higher-dimensional bulk spacetime, with
interactions of the standard model confined to the brane, while
gravitational interactions access the bulk. The original RS model
considered a Minkowski brane in an anti de Sitter (AdS) bulk, and
this was generalized to a Friedmann-Robertson-Walker (FRW) brane
in a Schwarzschild-AdS bulk~\cite{BDEL,cosmo}.

The massless 5-dimensional graviton has a massless (zero) mode
when projected onto the brane, but it also has a tower of massive
Kaluza-Klein (KK) modes, i.e. modes which have an effective mass
from a brane viewpoint. These massive modes introduce new features
into the generation and evolution of cosmological gravitational
waves, and the observational implications of these features
provide in principle constraints on the RS scenario. 
It has been
shown~\cite{LMW} (see also Refs.~\cite{GRS,FK,KKT}) that during
high-energy de Sitter inflation on the brane (in an AdS bulk), the
zero mode is generated with a scale-invariant spectrum, at an
amplitude which can be much greater than the corresponding general
relativity value:
\begin{eqnarray}
&& {\cal P}_T = \left[{8\over M_{P}^2}\left({H\over 2\pi}
\right)^{2}\right]
 \nonumber\\ &&{}\times\! \left\{ \sqrt{1+\ell^2H^2}+\ell^2
H^2\ln\left({ \ell H \over 1+ \sqrt{1+\ell^2H^2}
}\right)\!\right\}^{\!-1}
\end{eqnarray}
Here the general relativity expression is in square brackets, and
the braneworld modification is in curly brackets, with $\ell$ being the
curvature scale of the AdS bulk. (Table-top experiments to test
deviations from Newton's law currently impose the constraint
$\ell<0.1\,$mm.) At high energies, $\ell H\gg 1$, the correction
factor is large ($\to {3\over 2}\ell H$).

The massive modes are not excited during inflation~\cite{LMW}, and
the zero mode remains frozen after inflation while it is beyond
the Hubble radius. However, when the zero mode enters the Hubble
radius during radiation or matter domination, the separation
between the massless zero mode and massive bulk modes no longer
holds. A changing Hubble parameter induces mode-mixing, and an
initial zero mode will generate massive modes. An estimate of this
effect~\cite{GRS}, based on an instantaneous transition from a de
Sitter to a Minkowski brane, indicates that the effect will be
very small. This estimate has been refined by considering the
sharp transition to be from one de Sitter brane to another with
slightly smaller Hubble rate~\cite{KKT}.

In this paper we investigate the evolution of tensor metric
perturbations about an AdS bulk in Gaussian normal coordinates
defined with respect to an FRW brane. We first review the analytic
results for perturbations about the original RS solution with two
Minkowski branes embedded in AdS. We go on to develop a
perturbative analytic approximation (valid at low energies, near
the brane) to study the generation of massive modes with an FRW
cosmology on the brane.

\section{Gravitational wave equation and its exact solutions}

The generic 5D bulk metric that allows a spatially flat FRW metric
on the brane, can be written in Gaussian normal coordinates as
\begin{equation}
\label{metric} ^{(5)\!}d\bar{s}^2 = -N^2(t,y) dt^2 + A^2(t,y)
d\vec{x}\,^2 + dy^2\,,
\end{equation}
where the brane is located at $y=0$. Tensor metric perturbations
\cite{vdB,LMW,BMW,GRS,FK,KKT}
are given by
\begin{equation}
^{(5)\!}ds^2 = -N^2(t,y) dt^2 + A^2(t,y) \left[ \delta_{ij} +
h_{ij} \right] dx^i dx^j + dy^2\,,
\end{equation}
where $h_{ij}$ is 3-transverse ($\partial^ih_{ij}=0$) and
3-tracefree ($\delta^{ij}h_{ij}=0$). We will treat one Fourier
mode at a time, so that
\begin{equation}
h_{ij} = F(t,y)\, \hat{e}_{ij}(\bf{x}) \,,
\end{equation}
with $\hat{e}_{ij}$ a transverse and tracefree polarisation tensor
which is an eigenfunction of the spatial Laplacian
($\partial^k\partial_k\hat{e}_{ij}=-k^2\hat{e}_{ij}$).

The tensor amplitude $F$ obeys the 5D wave equation~\cite{LMW}
\begin{eqnarray}\label{w}
&& {1\over N^2} \left[ \ddot{F} + \left( 3{\dot{A}\over A}
 - {\dot{N}\over N} \right) \dot{F} \right] +
{k^2\over A^2} F = \nonumber\\ &&~~~~~~{} F'' + \left( 3{A'\over
A} + {N'\over N} \right) F' \,.
\end{eqnarray}

In the RS scenario, the bulk is an orbifold with Z$_2$-symmetry
about two fixed points, $y=0$ and $y=L$. This identifies points
$y\leftrightarrow-y$ and $y+L\leftrightarrow L-y$. We will assume
that we have branes at the two orbifold fixed points with just the
right surface energy-momentum tensors to satisfy the Israel
junction conditions. In the case of Minkowski branes, this simply
implies that the branes have constant brane tension. For a generic
FRW brane at $y=0$ this puts a strong physical
restriction~\cite{CGRT,BDL01} on the required equation of state on the
second FRW brane at $y=L$, which we will discuss in
Sec.~\ref{Sregulator}.

Perturbations must also satisfy the Z$_2$-symmetry. For a
Z$_2$-symmetric background this will be true, so long as the
initial conditions are Z$_2$-symmetric. The boundary conditions
for the metric perturbations at the branes, in the absence of
anisotropic pressure perturbations on the branes (which we will
assume) are
\begin{equation}
\label{boundary} F'|_{y=0}=0\,, \qquad F'|_{y=L}=0\,.
\end{equation}
This should be imposed on any initial conditions, and is then
preserved by the subsequent evolution.

In the special case when the FRW metric reduces to a static
Minkowski brane, the anti-de Sitter bulk metric Eq.~(\ref{metric}) is given by~\cite{RS}
\begin{equation}
N=A=\exp(-\mu y)\,,~~ \mu={1\over \ell}\,,
\end{equation}
where $\mu$ is the mass scale corresponding to the anti-de Sitter curvature scale
$\ell$. Here and in the following, the expressions hold for $0\leq
y \leq L$, and we consider the Z$_2$-symmetry as implicitly
imposed for other values of $y$. The wave equation~(\ref{w})
reduces to the simple separable form,
\begin{equation}
e^{2\mu y} \left( \ddot{F} + k^2 F \right) = F'' - 4\mu F'\,.
\end{equation}
Separating variables as
\begin{equation}\label{mode}
F(t,y) = \sum_m \varphi_m(t) F_m(y)\,,
\end{equation}
we find that
\begin{eqnarray}
\ddot{\varphi}_m + \left( m^2 + k^2 \right) \varphi_m &=& 0 \,,\\
F_m'' - 4\mu F_m' + m^2 e^{2\mu y}\,  F &=& 0 \,. \label{RSEm}
\end{eqnarray}

The general solutions are~\cite{RS}
\begin{eqnarray}
\varphi_m &=& c_+ e^{+i\sqrt{m^2+k^2}t} + c_-
e^{-i\sqrt{m^2+k^2}t} \,, \label{varphi}\\ F_0 &=& c_1+c_2e^{4\mu
y}\,,\\ F_m &=& e^{2\mu y} B_2 \left( {m\over\mu}\, e^{\mu y}
\right)~~~(m>0) \,,
\end{eqnarray}
where $c_\pm,c_1,c_2$ are constants and $B_2$ is a linear
combination of Bessel functions of order two. We only want the KK
modes that satisfy the boundary conditions~(\ref{boundary}). For
the zero mode, this requires
\begin{equation}
c_2=0\,.
\end{equation}
For the massive modes, the first condition (at $y=0$) requires
\begin{equation}
{m\over \mu} B_2'\left( {m\over \mu} \right) + 2 B_2\left( {m\over
\mu} \right) = 0 \,,
\end{equation}
where the prime here is the derivative with respect to the
argument of the Bessel function. This chooses the appropriate
linear combination of $J_2$ and $Y_2$. The second (at $y=L$)
requires
\begin{equation}
{m\over \mu}\, e^{\mu L}B_2'\left( {m\over \mu}\, e^{\mu L}
\right) + 2 B_2\left( {m\over \mu}\,e^{\mu L} \right) = 0 \, .
\end{equation}
These two conditions can be rewritten, using the recurrence
relation for Bessel functions $x B_n'+n B_n=x B_{n-1}$,  in the
simpler form
\begin{equation}
B_1\left( {m\over \mu} \right)=0=  B_1\left( {m\over \mu}\, e^{\mu
L}\right)\,,
\end{equation}
where $B_1$ is the same  linear combination as $B_2$ but with
$J_1$ and $Y_1$ instead of $J_2$ and $Y_2$.

The second condition can only be satisfied by particular values of
$m^2$, which selects a discrete set of KK eigenmodes in the
two-brane scenario. The allowed values of $m$ are the solutions of
\begin{equation}
J_1\left( {m\over \mu}\, e^{\mu L}\right) Y_1\left( {m\over \mu}
\right)- Y_1\left( {m\over \mu}\, e^{\mu L}\right) J_1\left(
{m\over \mu} \right) =0\,.
\end{equation}

The only case other than Minkowski branes for which the 5D wave
equation~(\ref{w}) is separable in Gaussian normal coordinates is
the case of a (anti-)de Sitter brane, which reduces to the
Minkowski brane as a limiting case. For completeness we give the
de Sitter mode functions in an appendix.

\section{Cosmological background solution}

The 5D bulk metric, Eq.~(\ref{metric}), for a general FRW brane at $y=0$ in an anti-de Sitter bulk is given explicitly by~\cite{BDEL}
\begin{eqnarray}
\label{aty} &&A(t,y) = a(t) \left\{ \cosh \mu y - [1+r(t)]\sinh
\mu y
\right\}\,,\\
\label{nty} &&N(t,y) = {A(t,y)\over a(t)}+3
\left[1+w(t)\right]r(t)\sinh \mu y\,,
\end{eqnarray}
where
\begin{equation}
r \equiv {\rho \over \lambda}\,,~ w\equiv {p\over \rho}\,,~
\mu={\kappa^2\lambda\over6}\,.
\end{equation}
Here $\lambda$ is the tension for a Minkowski brane, so that the high-energy regime is $\rho>\lambda$, and $\kappa^{2}=8\pi/M_5^3$, where $M_5$ is
the fundamental Planck scale. The effective Planck scale on the
brane is $M_P=\sqrt{M_5^3/\mu}$.

The two unknown functions of time, $a(t)$ and $r(t)$, are obtained
by solving the Friedmann and energy conservation equations on the brane,
\begin{eqnarray}
H= {\dot a \over a} &=& \sqrt{\mu^2 r(2+r)}\,, \\ \dot r &=& -3 H
(1+w) r\,.
\end{eqnarray}
For $r\ll 1$, one recovers the standard Friedmann equation. For
$r> 1$, the evolution is unconventional.

For a barotropic fluid with constant $w$ we have~\cite{BDEL}
\begin{eqnarray}
&& r\! = \left( {a_* \over a} \right)^{3(1+w)}\,, \\ &&
a^{3(1+w)}\! = a_*^{3(1+w)}\! \left[\! 1 + {3(1+w)\mu t \over 2}
\! \right]\! 3(1+w) \mu t,
\end{eqnarray}
where $a=a_*$ when $r=1$, so that $a_*$ marks the transition
between high- and low-energy regimes. In the case of radiation
domination, $w={1\over3}$, we have
\begin{eqnarray}
r &=& \left( {a_* \over a} \right)^4\,, \\ a &=& a_*
 \left( 1 + {2\mu t} \right)^{1/4} (4\mu t)^{1/4}\,, \label{arad}
\end{eqnarray}
recovering the standard evolution, $a\propto t^{1/2}$, for $t\gg
\mu^{-1}$.

The coordinate singularity $y_c$, defined by $A(t,y_c)=0$, and the
event horizon $y_h$, defined by $N(t,y_h)=0$, are given by
\begin{equation}\label{yc}
\coth\mu y_c=1+r\,,~~ \coth \mu y_h=1-\left(2+3w\right)r\,.
\end{equation}
There is no solution for $y_h$ if $w>-{2\over 3}$, whereas $y_c$
exists for all $\rho>0$ branes, and $y_c\to\infty$ as $r\to0$. For
$-1<w<-2/3$ we have $y_c<y_h$, while for a de Sitter brane
($w=-1$), we have $y_c=y_h$.

\subsection{Regulator brane}
\label{Sregulator}

To simplify the problem, and make it numerically tractable, we
will assume that there is a second regulator brane at a fixed
position $y_2=L < y_c$. Since $y_c(t)$ increases with the
expansion of the universe, the second brane remains within the
regular coordinate region. The horizon $y_h$ only exists for
$w<-{2\over3}$, which is the condition for inflation in the
high-energy limit~\cite{MWBH}.

The fixed position of the regulator brane constrains its energy
density and pressure via~\cite{BDL01}
\begin{eqnarray}
&&{\rho_L+\lambda_2\over \lambda} = {\sinh\mu L-(1+r)\cosh\mu
L\over \cosh\mu L-(1+r)\sinh\mu L}\,,\\ && {p_L-\lambda_2\over
\lambda} = -{2(\rho_L+\lambda_2)\over 3\lambda} \nonumber\\
&&~~{}-{1\over3}\left\{ {\sinh\mu L-[1-(2+3w)r]\cosh\mu L\over
\cosh\mu L-[1-(2+3w)r] \sinh\mu L}\right\}\,.
\end{eqnarray}
These imply
\begin{eqnarray}
&&(1+w_L)r_L=-(1+w)r \left[\cosh\mu L-(1+r)\sinh\mu L
\right]^{-1} \nonumber\\
&&~{}\times\left\{\cosh\mu L-[1-(2+3w)r] \sinh\mu
L\right\}^{-1}\,,
\end{eqnarray}
where $r_L=\rho_L/\lambda$, $w_L=p_L/\rho_L$. It follows that if
the physical brane is de Sitter ($w=-1$), then so is the regulator brane (with $w_{L}=-1$).
For $w>-{2\over3}$ we have the limiting cases,
\begin{eqnarray}
&&r\gg 1:\,  (1+w_L)r_L\! \approx\! {(1+w)\over [(2+3w)\sinh^2 \mu
L]r}\,,
\\ &&r\ll1 :\, (1+w_L)r_L\! \approx\! (1+w)re^{ \mu L}\,.
\end{eqnarray}
Thus the regulator brane requires $(1+w_{L})r_{L}\to0$ for both the very high energy regime on the physical brane $r\to\infty$, and the low energy regime on the physical brane $r\to0$.

\subsection{New coordinates}

It is convenient to introduce a new dimensionless variable for the
normal coordinate,
\begin{equation}\label{zc}
z\equiv e^{\mu y}\,,~ 1\leq z < z_c=\sqrt{1+{2\over r}}\,,
\end{equation}
where the range of validity follows from Eq.~(\ref{yc}). The
metric then reads
\begin{equation}
^{(5)\!}d\bar{s}^2 = -N^2(t,z) dt^2 + A^2(t,z) d\vec{x}\,^2 +
{dz^2\over \mu^2 z^2}\,,
\end{equation}
with Eqs.~(\ref{aty}) and~(\ref{nty}) being rewritten as
\begin{eqnarray}
\label{atz}
A(t,z)&=&{a(t)\over z}\left\{1 - {r(t)\over2} [z^2-1] \right\} \,, \\
\label{ntz} N(t,z)&=& {n(t)\over z}\left\{1 + [2+3w(t)]
{r(t)\over2} [z^2-1] \right\}\,.
\end{eqnarray}

The wave equation (\ref{w}) becomes
\begin{eqnarray}
\label{newwave} &&{1\over N^2} \left[ \ddot{F} + \left(
3{\dot{A}\over A} - {\dot{N}\over N} \right) \dot{F} \right] +
{k^2\over A^2} F = \nonumber\\ &&~{} \mu^2 \left[ z^2 F'' +
\left(z+ 3z^2{A'\over A} + z^2{N'\over N} \right) F'\right]\,,
\end{eqnarray}
where the prime from now on denotes $\partial/\partial z$.

\subsection{Causal propagation in the bulk}

The regulator brane will eventually have an effect on the physical
brane via gravitational wave propagation. Approximate and
numerical results will only be reliable up to the time when
perturbations from the regulator brane reach the physical brane.
The fastest signals in the bulk are zero-momentum ($k=0$) on the
brane. They follow null world-lines with (taking $n=1$)
\begin{equation}
{dz\over dt} = \pm \mu z n = \pm \mu \left[1 +
\left(1+{3w\over2}\right) r  (z^2-1) \right]\,,
\end{equation}
where, by Eq.~(\ref{zc}), $r(z^2-1)<2$. Hence for a FRW brane with
$w>-{2\over3}$, we have
\begin{equation}
\mu \leq \left| {dz\over dt} \right| < 3(1+w)\mu\,.
\end{equation}
In particular there is a lower limit on the time for causal
propagation from the regulator brane at $z=z_L=e^{\mu L}$ to our
FRW brane at $z=1$, given by
\begin{equation}
\Delta t_L > {e^{\mu L}-1 \over 3\mu (1+w)}\,.
\end{equation}

\section{Cosmological brane perturbations}

In general, the wave equation~(\ref{w}) for an FRW brane does not
separate in a Gaussian normal coordinate system where the physical
brane is fixed at $y=0$. The wave equation is only separable if
the background metric functions $A(t,z)$ and $N(t,z)$ are
themselves, and it can be shown that this is only possible in the
special case of constant brane surface density (which is the case
for a Minkowski brane~\cite{RSII} or de Sitter
branes~\cite{LMW,GS}).

Moreover, even if one could define a set of independent eigenmodes
for the bulk spacetime, the regulator brane in general represents
a time-dependent boundary condition. Thus the regulator brane,
moving relative to the normal coordinate system of the physical
brane, will also lead to mixing between modes. In what follows we
assume that the regulator brane remains at a fixed normal distance
to the physical brane. One would expect this to {\em
underestimate} the mode-mixing obtained when the second brane is
free to move.

The form of the bulk metric functions given in Eqs.~(\ref{atz})
and (\ref{ntz}) suggests that it should be possible to obtain an
approximate separable solution in the low-energy limit, $r
\rightarrow 0$, or close to the brane, $z\to1$, where $A$ and $N$
take the limiting forms
\begin{eqnarray}
A(t,z) \to {a(t) \over z} \,,\\
N(t,z) \to {n(t) \over z} \,.
\end{eqnarray}
In this limit we can study the limiting behaviour of $F$
analytically. (A similar approximation was independently discussed
in Ref.~\cite{MBV}.)

Formally we define the low-energy/near-brane regime by the
condition
\begin{equation}
\label{lenb} r(z^2-1) \ll 1 \,.
\end{equation}
If the regulator brane is fixed at finite $z_L<z_c$, then at
sufficiently late times (in an expanding cosmology with $w>-1$) we
will have $r\ll1/(z_L^2-1)$ and the asymptotic solution will
become a good approximation throughout the (finite) bulk.
Conversely, the low-energy/near-brane condition, Eq.~(\ref{lenb}),
will inevitably breakdown near the regulator brane in the limit
$z_L\to z_c$.

We can analytically estimate the effect of mode-mixing at finite
(but still small) $r$ by trying to build up a perturbative
solution starting from the low-energy/near-brane solutions,
presented in the next section, and then calculating the
corrections ${\cal O}(r)$, which we go on to do afterwards.

\subsection{Separable solution for low-energy/near-brane limit}

Using the canonical variable,
\begin{equation}
\label{tildeF} \tilde{F} = a z^{-3/2} F\,,
\end{equation}
and conformal time $\eta$ (so that $n=a$), we can write the wave
equation~(\ref{newwave}) for $r(z^2-1)\to0$ as
\begin{eqnarray}
\label{zerowavetildeE} {1\over a^2} {\cal D}_\eta \tilde{F}^{(0)} +
{k^2\over a^2}\tilde{F}^{(0)} - \mu^2 {\cal D}_z \tilde{F}^{(0)} = 0 \,,
\end{eqnarray}
where the second-order self-adjoint operators are
\begin{eqnarray}
\label{Oeta} {\cal D}_\eta &\equiv & {\partial^2 \over
\partial\eta^2} - {\ddot{a}\over
  a}\,, \\
\label{Oz} {\cal D}_z &\equiv & {\partial^2 \over \partial z^2} -
{15\over 4z^2}\,.
\end{eqnarray}
Here and from now on, a dot denotes $\partial/\partial\eta$.

The wave equation~(\ref{zerowavetildeE}) is separable and hence
its solution can be expressed in terms of KK eigenmodes,
\begin{equation}
\tilde{F}^{(0)}(\eta,z) = \sum_m \v_m^{(0)}(\eta) \psi_m(z)\,,
\end{equation}
where the mode functions are related to those in Eq.~(\ref{mode})
by $\v_m^{(0)} = a \varphi_m$ and $\psi_m = \N_m z^{-3/2} F_m$, with
${\cal N}_m$ a normalization constant. They obey the equations
\begin{eqnarray}
{1\over a^2} {\cal D}_\eta \v_m^{(0)} + {k^2\over
  a^2}\v_m^{(0)} = -m^2 \v_m^{(0)} \,,\label{vm} \\
{\cal D}_z \psi_m = -{m^2\over \mu^2} \psi_m \,.
\end{eqnarray}
The bulk eigenmodes obey the same equation as in the case of a
Minkowski brane, Eq.~(\ref{RSEm}), and hence we have (for $m>0$)
\begin{equation}
\psi_m = \N_m\sqrt{z} B_2\left({m \over \mu}\,z \right)\,,
\end{equation}
where the coefficient of normalization ensures that the $\psi_m$
constitute an orthonormal basis,
 \beq
\int_{1}^{{z_L}}dz\,\psi_m(z)\psi_{m'}(z)=\delta_{mm'} \,.
 \eeq

In the radiation era, $w={1\over 3}$, Eq.~(\ref{arad}) shows that
in the low-energy/ late-time regime,
\begin{equation}
a(\eta)= a_1\eta\,,~~a_1\equiv \sqrt{2}a_*^2\mu\,,
\end{equation}
and Eq.~(\ref{vm}) becomes
\begin{equation}  \label{phiRADIATION}
\ddot{v}_m^{(0)} +  \left(k^2 + m^2a_1^2\eta^2 \right) v_m^{(0)} = 0 \,.
\end{equation}
On large scales, or at late times, we can neglect the $k$-term,
and the solutions are:
\begin{eqnarray}
\label{v0} v_0^{(0)} &=& c_1\eta+c_2\,,\\ 
v_m^{(0)} &=& \eta^{1/2} B_{1/4}
\left({ma_1\over 2}\eta^2 \right)\,.
\end{eqnarray}
It follows that the massive modes $m\neq0$ decay on super-horizon
scales, unlike the massless mode $m=0$. For $\eta\gg 1$ and $k$
negligible, the massive modes behave as
\begin{equation}
v_m^{(0)} \approx \eta^{-1/2}\left[c_3 \cos\left({ma_1\over 2}\eta^2
\right) + c_4 \sin\left({ma_1\over 2}\eta^2 \right)\right].
\end{equation}

\subsection{Perturbative mode-mixing}

Using the bulk solution given in Eqs.~(\ref{atz}) and (\ref{ntz}),
we can now write the wave equation (\ref{newwave}) to first order
in $r$ as
\begin{eqnarray}
&&{1\over a^2} {\cal D}_\eta \tilde{F} + {k^2\over a^2}\tilde{F} -
\mu^2{\cal D}_z \tilde{F}=rS[\tilde F]\,, \\ && S[\tilde F] =
{(z^2-1)\over a^2} \left[ (2+3w){\cal D}_\eta \tilde{F}
\right.\nonumber\\ && \left.~~{}-
{3\over2}(1+w)(5+3c_s^2)\H(\tilde{F}^{\displaystyle\cdot}
-\H\tilde{F}) - k^2 \tilde{F} \right]\nonumber\\&&~~{} + (1-3w)
\left[ z\tilde{F}' + {3\over2} \tilde{F} \right] \,,
\end{eqnarray}
where ${\cal H}=\dot a/a=\partial_\eta a/a$ and $c_s^2=\dot p/\dot
\rho$. The full solution is a series expansion,
\begin{eqnarray}
\tilde{F} = \tilde{F}^{(0)} + \tilde{F}^{(1)} + \ldots \,,
\end{eqnarray}
where the zero-order solution $\tilde{F}^{(0)}$ is given by the
solution of Eq.~(\ref{zerowavetildeE}) and successive terms in
$\tilde{F}$ correspond to successively higher-order terms in $r$
in the wave equation~(\ref{newwave}). Higher-order corrections can
themselves be given as a sum over the zero-order bulk eigenmodes:
\begin{equation}\label{tfi}
\tilde{F}^{(i)}(\eta,z) = \sum_m \v_m^{(i)}(\eta) \psi_m(z)\,.
\end{equation}
Being eigenmodes of the self-adjoint operator ${\cal D}_z$ in
Eq.~(\ref{Oz}), the bulk modes $\psi_m$ form an orthonormal basis
for any function.

The first-order corrections are given by
\begin{eqnarray}
\label{fo} {1\over a^2} {\cal D}_\eta \tilde{F}^{(1)} + {k^2\over
a^2}\tilde{F}^{(1)} - \mu^2{\cal D}_z
\tilde{F}^{(1)}=rS[\tilde{F}^{(0)}].
\end{eqnarray}
Substituting the decomposition~(\ref{tfi}) and
projecting onto the basis of the functions $\psi_m$, leads to the
following equation that each of the coefficients
$\v^{(1)}_m(\eta)$ must satisfy:
\begin{eqnarray}
&&{1\over a^2} {\cal D}_\eta \v_m^{(1)} + {k^2\over a^2}\v_m^{(1)}
+m^2\v_m^{(1)} \nonumber\\ &&~{} = r \left\{\sum_n
\left(I_{mn}-\delta_{mn}\right) \left[ -(2+3w)n^2
\v_n^{(0)}\right. \right. \nonumber\\ &&\left.\left.~{} -
{3\over2a^2}(1+w)(5+3c_s^2)\H\left\{\dot{v}_n^{(0)}-\H\v_n^{(0)}
\right\}\right. \right. \nonumber\\ &&\left.\left.~{}
- 3(1+w){k^2\over a^2} \v_n^{(0)} \right] \right. \nonumber\\
&& \left.~{} + (1-3w)\sum_n \left(J_{mn}-{3\over
2}\delta_{mn}\right) \v_n^{(0)}
 \right\} \,.\label{vm1}
\end{eqnarray}
Here the matrix coefficients $I_{mn}$ and $J_{mn}$ are given
respectively by
\begin{eqnarray}
 I_{mn}&=&\int_1^{z_L}dz\, \psi_m(z) z^2
\psi_n(z)\,,\\
J_{mn}&=&\int_1^{z_L}dz\,  \psi_m(z) z{d\over dz} \psi_n(z)\,.
\end{eqnarray}
It is the presence of off-diagonal terms in the matrices $I$ and
$J$ that lead to mode mixing for $r>0$. Note that for $n=0$ we
have $\psi_0\propto z^{-3/2}$ and hence $J_{m0}=-3/2\delta_{m0}$.

For a regulator brane at fixed $z_L$, we expect mode-mixing to
occur mainly at early times (maximum $r$) and become small for
$r\ll 1/(z_L^2-1)$. Both analytic and numerical approaches are
thus limited to finite $z_L<z_c$. At a given time (fixed $r$) the
mixing becomes largest at large $z$. However the coordinate
singularity necessarily limits our analysis to
$z_L<z_c=\sqrt{1+2/r}$, and hence $r(z_L^2-1)<2$ in all cases,
suggesting a perturbative analysis should not be too bad in most
cases.

The zero-mode growing-mode solution for $k=0$ given in
Eq.~(\ref{v0}), has $\dot{v}_0^{(0)}={\cal H}v_0^{(0)}$ and hence
the whole right-hand-side of Eq.~(\ref{vm1}) vanishes even at
finite $r$. However an initial zero-mode configuration ($\v_m=0$
for all $m\neq0$) with finite $k$ evolves into a mixed mode
solution because the zero-order $\v_0^{(0)}\neq0$ acts as a source
term at first order for all $\v_m^{(1)}$ with $I_{m0}\neq0$.

We can see from Eq.~(\ref{vm1}) that the zero-mode with $n=0$
evolves independently of the massive modes only when $r=0$
(Minkowski brane) or $w=-1$ (de Sitter).

We have checked the approximation by numerical solution of the
full wave equation~(\ref{newwave}), in the case where
$w={1\over3}$ and only the lowest eigenmode ($m=0$) solution is
excited at lowest order, corresponding to an initial form of $F$
that is constant in $z$, and with $\dot{F}=0$. In this case all
the massive modes vanish initially.

\section{Discussion}

We have shown how in the low-energy/ near-brane limit we can
decompose bulk metric perturbations into RS modes which evolve
independently in the low-energy limit ($r\to0$). At finite $r>0$
and away from the brane, $z^2>1$, the bulk metric is not separable
in Gaussian normal coordinates defined with respect to a generic
FRW brane and this leads to mode mixing. At late times/low
energies, the bulk metric becomes separable for $r(z^2-1)\ll1$.

Our analytic approximation shows explicitly how mode-mixing occurs
when an initial massless mode, generated during inflation,
re-enters the Hubble horizon. The key result is Eq.~(\ref{vm1}).

On the physical brane at $z=1$, the 4D tensor metric perturbations, which are in principle constrained, e.g., by cosmic microwave background observations, have an amplitude
\begin{equation}
F|_{\rm brane}={1 \over a(\eta)}\left[ \tilde{F}^{(0)}(\eta,1) +
\tilde{F}^{(1)}(\eta,1) \right]\,,
\end{equation}
to lowest order in our approximation. The massive modes contribute
to the 4D tensor metric perturbations but their amplitude is
suppressed at the brane due to the RS volcano-type potential for $0<m^{2}<15\mu^{2}/4$~\cite{RSII}.
The massive modes contribute an anisotropic stress term in
the ``dark radiation'' term~\cite{BMW}
\begin{equation}
\delta E_j^i = - \frac12 \left(h_j^i\right)'' - \frac{A'}{A}\left(
h_j^i\right)' \,,
\end{equation}
where the derivatives here are with respect to $y$.

We note that any analytic or numerical analysis based on Gaussian normal
coordinates faces some significant practical limitations that may
mean that it is not particularly well-suited for more detailed
calculations. In particular, the existence of a coordinate
singularity at $z=z_c$, a finite proper distance from the brane,
makes it impossible to treat an infinite bulk. Introducing a
second brane at a fixed Gaussian normal distance $z=z_L<z_c$
may require an unphysical equation of state on the second brane. Thus
unphysical effects may propagate to the ``physical'' (e.g.,
radiation-dominated) brane in a finite time.

~\\ {\bf Acknowledgements}

A visit by DL to Portsmouth to initiate this work was supported
by a PPARC grant PPA/G/S/1999/00138. 
RM is supported by PPARC. DW is supported by the Royal Society.
This research was supported in part by the National Science Foundation under Grant No.~PHY99-07949.

~\\ {\em Note added:}

In the final stages of this work, Ref.~\cite{hkt} appeared,
dealing with the same topic.

\appendix

\section{DE SITTER BRANE}

The bulk background metric is given by
 \beq
^{(5)\!}d\bar{s}^2=N^2(y)\left[-dt^2+e^{2H_0
t}\delta_{ij}dx^idx^j\right]+ dy^2\,,
 \eeq
where $a=e^{H_0t}$, $H_0$ is constant and $N(y)$ is given by
 \beq
N(y)= \cosh \mu y - (1+r)\sinh \mu y\, ,
 \eeq
with $r$ constant. The wave equation has the form
 \beq
\ddot F+3H_0 \dot F+{k^2\over a^2}F=N^2 F''+4NN' F'\,,
 \eeq
and is separable, so that $F(t,y)$ can be written in the form of
Eq.~(\ref{mode}), and
\begin{eqnarray}
\ddot{\varphi}_m +3H_0\dot{\varphi}_m+
  \left( m^2 + {k^2\over a^2} \right) \varphi_m &=& 0 \,,\\
F_m'' + 4{N'\over N} F_m' +  {m^2\over N^2} F_m &=& 0 \,.
\end{eqnarray}
The general solutions are given by
 \beq
{\varphi}_m(t)=\exp\left(-{3\over 2}H_0 t\right)
B_\nu\left({ke^{-H_0t}\over H_0}\right)\,,
 \eeq
with
 \beq\label{nu}
\nu^2={9\over 4}-{m^2\over H_0^2}\,,
 \eeq
and
 \beq
F_m(y)={A^\nu_{3/2}\left(\sqrt{1+\mu^2N^2/ H_0^2}\right)\over
N^{3/2}}\,,
 \eeq
where $A^\nu_{3/2}$ is a linear combination of associated Legendre
functions. Equation~(\ref{nu}) indicates the existence of a mass
gap~\cite{GS,LMW,FK} between the zero mode and the start of the
massive KK tower at $m={3\over 2}H_0$.

\end{document}